\documentstyle[aps,epsf]{revtex}
\twocolumn

\newcommand{\infig}[2]{\begin{center}\mbox{\epsfxsize #2
\epsfbox{#1}}\end{center}}

\begin{document}
\input {epsf}

\newcommand{\beq}{\begin{equation}}
\newcommand{\eeq}{\end{equation}}
\newcommand{\beqa}{\begin{eqnarray}}
\newcommand{\eeqa}{\end{eqnarray}}

\def\ov{\overline}
\def\onlyif{\rightarrow}

\def\openone{\leavevmode\hbox{\small1\kern-3.8pt\normalsize1}}

\def\a{\alpha}
\def\b{\beta}
\def\g{\gamma}
\def\r{\rho}
\def\minus{\,-\,}
\def\eks{\bf x}
\def\kay{\bf k}

\def\ket#1{|\,#1\,\rangle}
\def\bra#1{\langle\, #1\,|}
\def\braket#1#2{\langle\, #1\,|\,#2\,\rangle}
\def\proj#1#2{\ket{#1}\bra{#2}}
\def\expect#1{\langle\, #1\, \rangle}
\def\trialexpect#1{\expect#1_{\rm trial}}
\def\ensemblexpect#1{\expect#1_{\rm ensemble}}
\def\kpsi{\ket{\psi}}
\def\kphi{\ket{\phi}}
\def\bpsi{\bra{\psi}}
\def\bphi{\bra{\phi}}

\def\ditto{\rule[0.5ex]{2cm}{.4pt}\enspace}
\def\th{\thinspace}
\def\ni{\noindent}
\def\thirty{\hbox to \hsize{\hfill\rule[5pt]{2.5cm}{0.5pt}\hfill}}

\def\set#1{\{ #1\}}
\def\setbuilder#1#2{\{ #1:\; #2\}}
\def\Prob#1{{\rm Prob}(#1)}
\def\pair#1#2{\langle #1,#2\rangle}
\def\Id{\bf 1}

\def\dee#1#2{\frac{\partial #1}{\partial #2}}
\def\deetwo#1#2{\frac{\partial\,^2 #1}{\partial #2^2}}
\def\deethree#1#2{\frac{\partial\,^3 #1}{\partial #2^3}}

\newcommand{\xx}{{\scriptstyle -}\hspace{-.5pt}x}
\newcommand{\yy}{{\scriptstyle -}\hspace{-.5pt}y}
\newcommand{\zz}{{\scriptstyle -}\hspace{-.5pt}z}
\newcommand{\kk}{{\scriptstyle -}\hspace{-.5pt}k}
\newcommand{\sx}{{\scriptscriptstyle -}\hspace{-.5pt}x}
\newcommand{\sy}{{\scriptscriptstyle -}\hspace{-.5pt}y}
\newcommand{\sz}{{\scriptscriptstyle -}\hspace{-.5pt}z}
\newcommand{\sk}{{\scriptscriptstyle -}\hspace{-.5pt}k}

\def\openone{\leavevmode\hbox{\small1\kern-3.8pt\normalsize1}}

\title{Quantum secret sharing using pseudo-GHZ states}
\author{
W. Tittel, H. Zbinden, and N. Gisin
\\
\small
{\it Group of Applied Physics, University of Geneva, CH-1211, Geneva 4,
Switzerland}}
\maketitle

\abstract{We present a setup for quantum secret sharing using 
pseudo-GHZ states based
on energy-time entanglement. In opposition to true GHZ states, 
our states do not
enable GHZ-type tests of nonlocality, however, they bare the same 
quantum correlations. 
The relatively high coincidence count rates found in our setup
enable for the first time an application of a quantum 
communication protocoll
based on more than two qubits.} 

\vspace{0.5 cm}
\normalsize
  
%-----------------------------------------------------------------------
%                          Introduction
%-----------------------------------------------------------------------
Entangled particles play the major role both as candidates for tests of
fundamental physics \cite{Bell,Aspect,TittelBell,WeihsBell} as well as 
in the whole field of quantum communication \cite{quantumcommunication}. 
Until recently, most work 
has been focussed on two-particle correlations.  
Since a couple of years, however, the interest in 
multi-particle entanglement 
-- which we identify in this article with n$>$2 --
is growing rapidly. 
From the fundamental side, particles in so-called GHZ states enable 
new tests of 
nonlocality \cite{GHZ}. 
From the side of quantum communication, more and more ideas 
for applications, 
like quantum secret sharing \cite{Hillery,Karlsson}, 
emerge. A major problem still is 
the lack of multi-photon sources. Nonlinear effects that enable 
to "split" a pump photon
into more than two entangled photons are extremely low efficient, 
and experiments still lie in the future. Recently Bouwmeester et 
al. could demonstrate 
a different approach where they started with two pairs of entangled 
photons and 
transformed them 
via a clever measurement into three photons in a GHZ state and a 
fourth independant 
trigger photon \cite{expGHZ}.
In this letter we present another method to create what we term 
pseudo-GHZ states. 
It is based on a recently developed novel source for 
quantum communication, 
creating entangled photons in energy-time Bell-states 
\cite{newsource,energytimecrypto}.
In opposition to "true" GHZ states, the three photons being in 
the pseudo-GHZ state do not consist of 
three downconverted photons but only of two downconverted ones plus 
the pump photon. 
We will comment on similarities and differences compared to true GHZ 
states and
demonstrate a first application of our states for quantum secret sharing. 
In this case, the difference to true GHZ states is not 
only of no importance, but enables for the first time 
to realize a multi-particle application of quantum communication. 

%-------------------------------------------------------------------
%                                GHZ states
%--------------------------------------------------------------------

Entangled states of more than two qubits, so-called GHZ states, can 
be described in the form
\begin{equation}
\ket{\psi}_{GHZ}=\frac{1}{\sqrt{2}}\bigg(\ket{0}_1\ket{0}_2\ket{0}_3+
\ket{1}_1\ket{1}_2\ket{1}_3\bigg)
\label{GHZ}
\end{equation}
\noindent
where $\ket{0}$ and $\ket{1}$ are orthogonal states in an arbitrary
Hilbert space and the indices label the
particles (in this case three). As shown by Greenberger, 
Horne and Zeilinger in 1989
\cite{GHZ}, 
the attempt to find a 
local model able to reproduce the quantum correlations
faces an inconsistency. In the multi-particle case, the 
contradiction occurs 
already when trying to 
describe the perfect correlations. 
Thus, demonstrating these correlations
directly shows that nature can not be described by local theories.
However, since it will never be possible to experimentally 
demonstrate perfect correlations, 
the question arises whether there is some kind of threshold, similar 
to the one
given by Bell inequalities for two-particle correlations \cite{Bell}, 
that enables to seperate
the "non-local" from the "local" region. 
Indeed, the generalized Bell inequality for the three-particle case
\cite{Bell3} 

\centerline{
$S_3^{\lambda}=\bigg|E(\a',\b,\g)+E(\a,\b',\g)+$}
\begin{equation}
+E(\a,\b,\g')-E(\a',\b',\g')\bigg|\leq2
\label{S3}
\end{equation}
\noindent
with E($\a$,$\b$,$\g$) the expectation value for 
a correlation measurement with analyzer settings $\a,\b,\g$
can be violated by quantum mechanics, the maximal value being
\begin{equation}
S_3^{qm} = 4.
\end{equation}
\noindent 
For instance, finding a correlation function of the form
$E(\a,\b,\g)=Vcos(\a+\b+\g)$
with visibility V above 50\% shows that the correlations under test
can not be described by a local theory.
Note that this
value is much lower than in the two-particle case where the threshold 
visibility is 
$\approx$ 71 \%.

----------------------------------------------------------------------
%                        Quantum secret sharing
%---------------------------------------------------------------------

Quantum secret sharing \cite{Hillery,Karlsson} is an expansion of 
the "traditional" quantum key distribution to more than two parties. 
In this new application of 
quantum communication, a sender, usually called Alice, distributes a 
secret 
key to two other parties, Bob and Charly, in a way that 
neither Bob nor Charly alone have any information about the key,
but that together they have full information.
Moreover, an eavesdropper trying to get some information
about the key creates errors in the transmission data and thus 
reveals his
presence. The motivation for secret sharing is to guarantee that Bob and 
Charly must cooperate -- one of them might be dishonest --
in order to do some task, one might think for instance of accessing 
classified information. 
As pointed out by Hillery et al. \cite{Hillery}, this protocol can 
be realized
using GHZ states.
Assume three photons
in a GHZ state of the form (\ref{GHZ}) 
%which propagate towards Alice, Bob and Charly, respectively, 
with $\ket{0}$ and $\ket{1}$ being 
different modes of the particles (Fig. 1). 
After combining the modes at beamsplitters located at Alice's, Bob's and 
Charly's, respectively,
the probability to find the three photons in any combination of 
outputs ports 
depends on the settings 
$\a,\b,\g$ of the phase shifters:
\begin{equation}
P_{i,j,k}=\frac{1}{2}\big(1+ijk~cos(\a+\b+\g)\big)
\label{probability}
\end{equation}
\noindent
with $i,j,k=\pm1$ labeling the different output ports.
Before every measurement, Alice chooses randomly one out of two 
phase values
($0,\pi/2$), Bob applies a phase shift of either $-\pi/2$ or 0,
and Charly chooses between $\pi/2$ and $3\pi/2$. 
After a sufficient number of runs, they publicly 
identify the cases where all detected a photon.
All three then announce the phases chosen and single out the cases where
the sum adds up either to 0 or to $\pi$.
Note that the probability function (Eq. \ref{probability}) yields 1 
for these cases.
Denoting l=cos($\a'$+$\b$+$\g$)=$\pm$1 and using $P_{i,j,k}$=1, 
Eq. \ref{probability} leads to

\begin{equation}
ijkl=1.
\end{equation}

\noindent
At this point, each of them knows two out of the values $i,j,k,l$. 
If now Bob and Charly
get together and join their knowledge, they know three of the four 
parameters and 
can thus determine the last one, which is also known to Alice. 
Identifying "--1" with bitvalue "0" and "+1" with "1", the 
correlated sequences of parameter values 
can then be turned into a secret key. Note that this scheme 
is completely symmetric. Any of the three can force the two other to 
collaborate
in order to get information about his key, 
which in turn enables to read his confidential message.   
Like in two-party quantum cryptography, 
the security of quantum secret sharing using GHZ states is given by the
fact that the measurements are made in noncommuting bases 
\cite{Hillery,Karlsson,eavesdropping}. 
An eavesdropper, including a dishonest Alice, Bob or Charly, is thus 
forced to guess about the bases that will be 
chosen. The
fact that she will guess wrong in half of the cases then leads to 
detectable errors
in the transmission data which reveal her presence. 

\begin{figure}
\infig{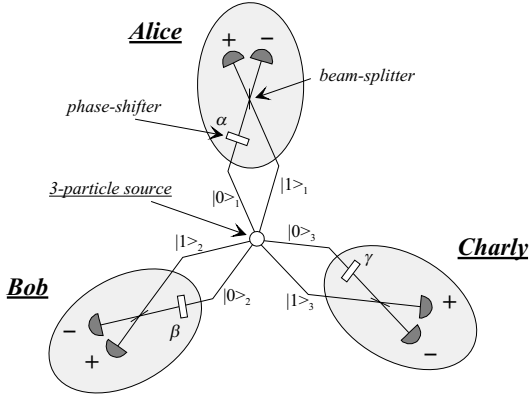}{0.80\columnwidth}
\caption{Schematics for quantum secret sharing using GHZ states.}
\label{idea} 
\end{figure}%\noindent

\begin{figure}
\infig{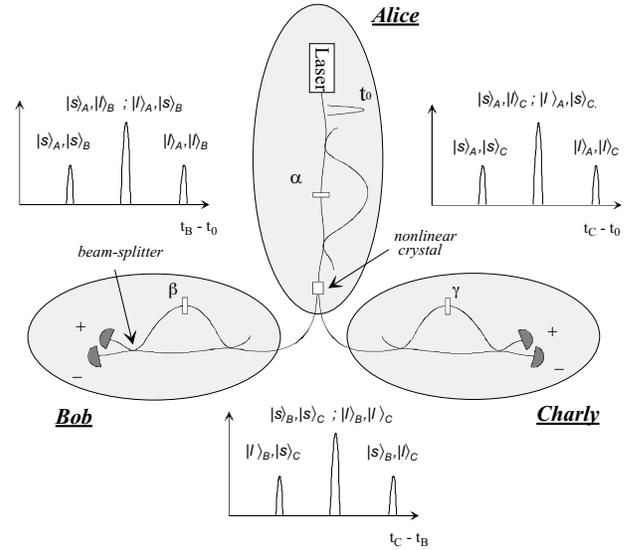}{0.95\columnwidth}
\caption{Principle setup for quantum secret sharing using 
energy-time entangled
pseudo-GHZ states. Here shown is a fiberoptical realization.}
\label{setup} 
\end{figure}%\noindent

We now explain how to implement secret sharing using our source 
(see Fig. 2). 
A short light pulse emitted at time $t_0$ enters an interferometer 
having a 
path length difference which is large compared to the duration of 
the pulse.
The pulse is thus split into two pulses of smaller amplitude, 
following each other with a fixed phase relation. 
The light is then 
focussed into a nonlinear crystal where some of the pump photons are 
downconverted into photon pairs.
The pump energy is assumed to 
be such that the possibility to create more than one pair from
one initial pump pulse can be neglected. %\cite{comparefaintpulses}.
This first part of the setup is located at Alice's.
The downconverted photons are then separated and send to Bob and 
Charly, respectively. 
Both of them are in possession of a similar interferometer as Alice, 
introducing 
exactly the same difference of travel times. 
We assume the transmission probabilities
via the different arms of any of the three interferometers to be alike.
The two possibilities for the photons to pass through any device lead 
to three 
time differences between emission of the pump pulse at Alice's and 
detection of signal
or idler photon, as well as between the detection of one downconverted
photon at Bob's and the correlated one at Charly's (Fig. 2).
Looking for example 
at the possible time differences between detection at Bob's 
and emission of the pump pulse ($t_B-t_0$), we find three different 
terms. The 
first one is due to "pump pulse travelled via the short arm and Bob's 
photon travelled 
via the short arm" to which we refer as $\ket{s}_A;\ket{s}_B$. Please 
note that this
notation considers the pump pulse as being a single photon (now termed 
"Alice's photon"), stressing the fact that only one pump photon is 
anihilated to 
create one photon pair. Moreover, the fact that this state 
is not a product state is taken into account by separating the two kets 
by ";". 
The second time difference is either due 
to $\ket{s}_A;\ket{l}_B$, or to $\ket{l}_A;\ket{s}_B$, 
and the third one to $\ket{l}_A;\ket{l}_B$. Similar time spectra arise 
when looking at the 
time differences between emission at Alice's and detection at Charly's 
($t_C-t_0$), 
as well as between
the detections at Bob's and Charly's ($t_C-t_B$).
Selecting now only processes, leading to the central peaks 
\cite{postselection}, we find two
possibilities:
Either Alice's photon traveled via the long arm and Bob's as well as 
Charly's took
the short ones, or Alice's photon choose the short arm and Bob's and 
Charly's both the 
long ones. If both possibilities are indistinguishable, the process is
described by
\begin{equation}
\ket{\psi}=\frac{1}{\sqrt{2}}\bigg(\ket{l}_A;\ket{s}_B\ket{s}_C +
e^{i(\a+\b+\g)} \ket{s}_A;\ket{l}_B\ket{l}_C\bigg),
\label{pseudoGHZ}
\end{equation}
with phases $\a,\b,\g$ in the different interferometers. 
The maximally entangled state (\ref{pseudoGHZ}) is 
similar to the GHZ state given in Eq.\ref{GHZ}, the difference 
being that the 
three photons do not exist at the same time (remember the ";"). 
Therefore,
our state is obviously of no significance concerning GHZ-type tests 
of nonlocality. 
To stress this difference, we call it pseudo-GHZ state. However, 
the probability-
function describing the triple coincidences 
(Eq.\ref{probability}) -- in our case
between emission of a pump pulse and detection at Bob's and 
Charly's -- is the same 
as the one originating from a true GHZ state.
To avoid the complication of switching the pump laser 
randomly between one of the two input ports -- equivalent
to detecting a photon in one or the other output port --, we let Alice
chose between one of four phase values $\a'$ (0,$\pi/$2,$\pi$,3$\pi$/2).
To map the choice of phases to the initial scheme where the information 
of Alice, Bob and Charly
is given by a phase setting and a detector label, we assign a different 
notation to 
characterize Alice phases (table 1).
Using this convention, we can implement the same protocol as given above, 
the advantage being
the fact that our setup circumvents creation and coincidence detection 
of photon triples.
The emission of the 
bright pump pulse can be considered as detection of a photon with 100 \%
efficiency. Moreover, no low efficient triple photon generation 
is necessary. 
This leads to much higher triple coincidence rates, 
enabling the demonstration
of a multi-photon applications of quantum communication.
One might question the security of our setup, 
the weak point being the channel leading from Alice's interferometer 
to the crystal. Here,
the light is classical and the phase could be measured without 
modifying the system.
However, since this part is controlled by Alice and 
the parts, physically accessible to an eavesdropper 
carry only quantum systems, our realization does not incorporates 
any loophole.

%----------------------------------------------------------------------
%               experimental realization.
%-----------------------------------------------------------------------
The experimental setup is described in 
\cite{energytimecrypto}, where it is used to demonstrate
two-party quantum key distribution 
using energy-time Bell states.
We will thus give only a brief outline.
To generate the short pump pulses, we use a pulsed diode 
laser (PicoQuant PDL 800), 
emitting 600ps (FWHM) pulses of 
655 nm wavelength at 
a repetition frequency of 80 MHz. The light is channeled through a 
fiberoptical 
Michelson interferometer (path length difference corresponding to 1.2 ns 
travel time difference) and focussed into a 4x3x12 mm $KNbO_3$ crystal,
producing photon pairs at 1310 nm wavelength. 
The average power 
before the crystal is $\approx$ 1 mW, and the energy per 
pulse -- remember that each initial pump pulse is now split into two --
$\approx$ 6 pJ. 
After absorption of the red pump light, the downconverted photons are 
separated and are guided 
to fiberoptical Michelson interferometers, 
located at Bob's and Charly's, respectively. To access the second 
output port, usually coinciding with the input port for this kind 
of interferometer,
we implement 3-port optical circulators. The interferometers 
incorporate equal
path length differences, and the travel time difference is 
the same than the one
introduced by the interferometer acting on the pump pulse. 
The output ports
are connected to single-photon counters -- passively quenched 
germanium avalanche photodiodes, operated in
Geiger-mode and cooled to 77 K. We operate them at dark count rates 
of 30 kHz, leading to quantum efficiencies of $\approx$ 5 \% 
and single photon detection rates of 4-7 kHz. 
The electrical output from each detector is fed into a fast AND-gate,
together with a signal, coincident with 
the emission of a pump pulse. 
We condition the detection at Bob's and Charly's on the central peaks
($\ket{s}_P;\ket{l}_A$ and $\ket{l}_P;\ket{s}_A$, 
and $\ket{s}_P;\ket{l}_B$ and $\ket{l}_P;\ket{s}_B$, respectively).
Looking at coincident detections between two AND-gates -- equivalent to 
triple coincidences --,
we finally select only the interfering processes for detection. 

\begin{figure}
\infig{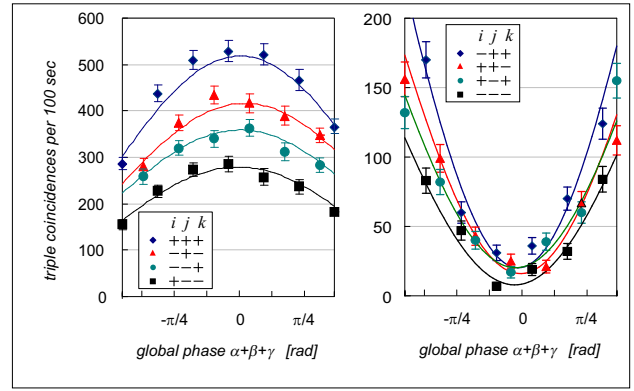}{0.95\columnwidth}
\caption{
Result of the measurement when changing the phase in Alice 
interferometer. 
The different mean values are due to non-equal quantum efficiencies 
of the single
photon detectors.
The values for the case $i$="-1" have 
been found corresponding to table I.
For global phase zero, we find a mean QBER of (3.9$\pm$0.4)\%.
If Bob and Charly both detect a photon in the "+"-labeled detectors 
in this case ($l$=+1), they know that Alice value $i$ must be +1 as well.
}
\label{results} 
\end{figure}

Since the stability of our interferometers is not sufficient to maintain 
stable phases over a long time, we demonstrate that our source can
be used for quantum secret sharing by continuously changing 
the phases in Alice's as well as in 
Bob's interferometer. We observe sinusoidal
fringes in the triple coincidence rates with 
maximum count rates around 1600
in 100 sec and minimum ones around 70 (see Fig. 3). 
Visibilities are inbetween 89.3
and 94.5\% for the different detector combinations,
leading to a mean visibility of 92.2$\pm$0.8\% 
and a quantum bit error rate BER --
the ratio of errors to detected events -- of (3.9$\pm$0.4)\%. 
The critical visibility above which the 
information that might have been obtained by an eavesdropper
can be made arbitrarily small using classical error correction and
privacy amplification is not known yet.
In case of two-party quantum
key distribution, it corresponds exactly to 
a violation of two-particle Bell inequalities \cite{eavesdropping}. 
It is thus reasonable to compare the mean 
visibilitiy to the value given by generalized Bell 
inequality (Eq.\ref{S3}), even if our setup 
does not incorporate GHZ-type nonlocality: 
The found visibility of 92.2$\pm$0.8\% 
is more than 50 standard deviations ($\sigma$) 
higher than the the threshold visibility for the three-particle case. 
Moreover, it is more than 25 $\sigma$ above 71\%, 
the value given by standard (two-particle) Bell inequalities. 
Within this respect, it is also interesting
to calculate $S_{exp}$: 
We find $S_{exp}$=3.69, well above $S^{\lambda}_3$=2 (Eq.\ref{S3}).
Therefore, the performance of our source is good enough to detect
any eavesdropping and to ensure secure key distribution. Moreover, the
bit-rate of $\approx$ 16 Hz underlines its potential for real applications.
To compare our coincidence rate to an experiment using true GHZ 
states \cite{expGHZ}, 
Bouwmeester et al. found one GHZ-state per 150 sec.
However, in order to really implement our setup for quantum secret sharing, 
an active phase stabilization and a fast switch still have to be 
incorporated \cite{stability}.

Like in all experimental quantum key distribution, the QBER is non-zero, 
even in the 
absence of any eavesdropping. The observed 4\% 
can be traced back to wrong counts from
accidentally correlated event at the single-photon counters,
non-perfect localization of the pump pulse, limited resolution of 
the single-photon
detectors and non-perfect interference. Note that the number of 
errors due to the last 
mentioned points decrease with distance (caused by higher 
transmission losses)
and thus do not engender an increase of the QBER. 
In opposition, the number of errors due to accidental coincidences 
stays almost 
constant since it is mostly due to detector noise. However,  
it causes only 10\% 
of the total errors in our laboratory demonstration. Therefore, the QBER
will increase only at a small rate,
enabling quantum secret sharing over tens of kilometers.

In conclusion, we demonstrated quantum secret sharing using 
energy-time entangled
pseudo-GHZ states in a laboratory experiment. 
We found bit-rates of around 16 Hz and quantum bit error rates of 4\%, 
low enough to ensure secure key distribution. 
The advantage of our scheme is the fact that neither triple-photon 
generation nor 
coincidence detection
of three photons is necessary, enabling for the first time
an application of a multi-particle quantum communication protocol. 
Moreover, since energy time entanglement can be 
preserved over long distances \cite{TittelBell}, our results are very 
encouraging for
realizations of quantum secret sharing over tens of kilometers.

%---------------------------------------------------------------------
%                               acknowledgements
%---------------------------------------------------------------------

We would like to thank J.-D. Gautier for technical support 
and Picoquant for fast delivery of the laser.
Support by the Swiss FNRS and the European QuComm IST project is 
gratefully 
acknowledged.

%--------------------------------------------------------------------

%---------------------------------------------------------------------

\begin{table}
\begin{tabular}[5]{c|cccc}
$\a'$&0&$\pi/2$&$\pi$&$3\pi/2$\\
\hline
$\a$&0&$\pi/2$&0&$\pi/2$\\
$i$&1&1&-1&-1\\
\end{tabular}
\caption{Mapping of the four possible phases $\a'$ at Alice's to two 
phase values $\a$ and 
the parameter $i$.}
\end{table}
\noindent

\begin{thebibliography}{99}
\bibitem{postselection} The desired processes are thus postselected, 
        leading to an
        decrease of count rates. It is in principle possible to 
        create only the 
        events in the central peaks using switches instead
        of passive couplers \cite{newsource}.
\end{thebibliography}
\end{document}